\documentclass[journal]{IEEEtran}
\usepackage{cite}
\usepackage{geometry, graphicx, subcaption}
\newsavebox{\largestimage}
\ifCLASSINFOpdf
\else
\fi
\usepackage[cmex10]{amsmath}
\usepackage{algorithm,algorithmic,mathtools,commath}
\usepackage{fullpage}
\usepackage{times}
\usepackage{fancyhdr,graphicx,amsmath,amssymb}
\usepackage{array}
\usepackage{url}
\usepackage{wasysym}
\usepackage{caption}
\usepackage{subcaption}
\usepackage{amsmath}
\usepackage{romannum}
\usepackage{textcomp}
\usepackage{color,soul}
\usepackage{xcolor}
\usepackage[framemethod=tikz]{mdframed}
\usepackage{lipsum}
\usepackage{multirow, diagbox, hhline, booktabs}
\colorlet{shadecolor}{yellow!20}
\colorlet{soulyellow}{yellow!20}
\sethlcolor{soulyellow}

\begin{document}

\title{Dynamic Control of a Fiber Manufacturing Process using Deep Reinforcement Learning}

\author{Sangwoon Kim*, David~Donghyun~Kim*
        and
        Brian W. Anthony
        
\thanks{*S. Kim and D. D. Kim has contributed equally to this paper}
\thanks{The authors are with the Department of Mechanical Engineering, MIT, Cambridge, MA, 02139 USA. Corresponding author: Brian Anthony (banthony@mit.edu, 617-324-7437).}
}
\maketitle

\begin{abstract}
This paper presents a model-free deep reinforcement learning (DRL) approach for controlling a fiber drawing system. The custom DRL-based control system predictively regulates fiber diameter and produces a fiber with a desired, constant or non-constant, diameter trajectory, i.e. diameter variation along the fiber length.  Physical models of the system are not used. The system was trained and tested on a compact fiber drawing system, which has non-linear delayed dynamics and stochastic behaviors. For a reference trajectory with random step changes, after 1 hour of training, the DRL controller showed the same root mean squared error (RMSE) as an optimized PI controller; after 3 hours of training, it achieved the performance of a quadratic dynamic matrix controller (QDMC). While the PI feedback controller showed 3.5 seconds of time lag in a step response, the DRL controller showed less than a second of time lag. Controller performance tests on trajectories not used in the training process are conducted; for a sine sweep reference trajectory, the DRL controller maintained an RMSE under 40 $\mu m$ up to a frequency of 45 mHz, compared to 25 mHz for QDMC.
\end{abstract}

\begin{IEEEkeywords}
Fiber fabrication, Learning control systems, Process control, Neural network applications
\end{IEEEkeywords}

\IEEEpeerreviewmaketitle

\begin{figure}[t]
\centering
\includegraphics[width=0.9\linewidth]{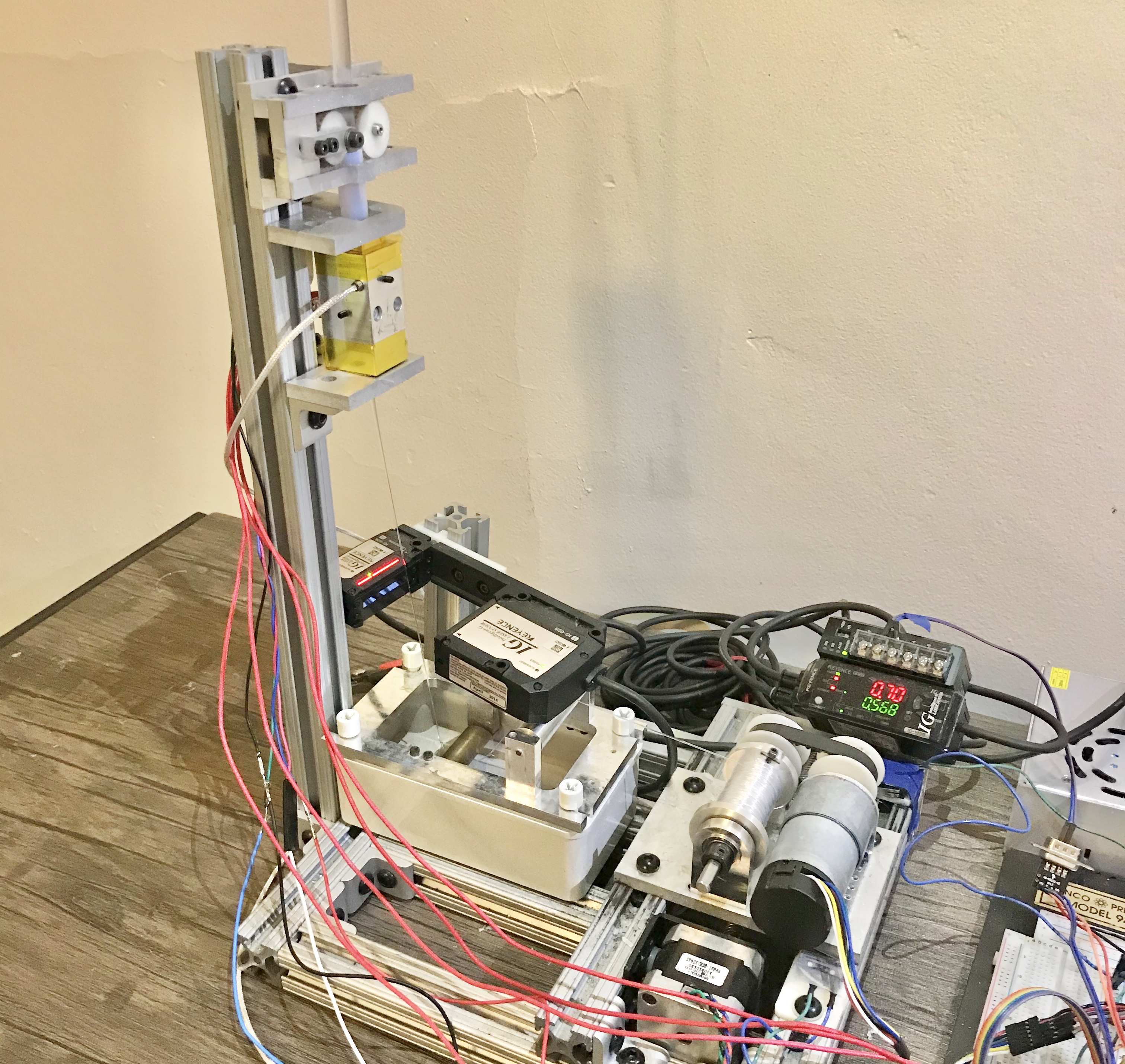}
\caption{Compact fiber drawing system}
\label{desktop}
\end{figure}

\section{Introduction}

\IEEEPARstart 
{O}{ptical} fiber is an integral part of communication technology.  New fiber designs may extend the functionality of fibers beyond communication into sensing, i.e. smart fibers. Microstructures \cite{Xue:05}  and semi-conducting materials inside an optical core, when thermally drawn, will create fiber-based sensors \cite{Abouraddy2007}. Thermally drawn fibers are used as waveguides for interaction with neurons\cite{ParkNature}.  Hollow polymer fiber is used to grow nerve cells for repairing severed nerves \cite{KOPPES201627}.  New fiber designs, for new applications, require innovations in the machines, processes, and controllers used to produce fiber. Fiber manufacturing systems used for research, and fast prototyping, require control systems capable of supporting new geometries, new materials, and new fabrication methodologies.

The optical fiber drawing process heats a large diameter glass rod (preform), which is then pulled axially from the furnace to generate a thin fiber. Control methods for fiber diameter regulation have been studied. Mulpur and Thompson developed a modal diameter control method based on simulation\cite{MulpurDiameterControl}; they also developed non-linear control strategies\cite{MulpurNonLinear}. State-space modeling of the optical fiber drawing process coupled with Linear Quadratic Gaussian optimal controllers was investigated \cite{TchikandaModel, TchikandaControl}.  Models of the neck-down profile and control of the draw tension were used to enable high-speed production \cite{WeiFreeSurface}. This long history of modeling and controlling the optical fiber manufacturing process focuses on maintaining the diameter at a fixed set point. When setpoints change, new state models operating at the new setpoints are required \cite{TchikandaModel}; a dynamic model for the controlled transition between different set points is also required. In this paper, we use deep reinforcement learning (DRL) to teach a controller how to operate at any setpoint, of diameter, and how to track a dynamically varying diameter trajectory.

The growth in availability of computation resources enables the development of control methods that utilize machine learning, especially DRL. DRL based algorithms outperform humans in playing the game of Go \cite{Silver2016} and Atari \cite{Mnih2013}. It is successfully used in the simulation of physical tasks \cite{Lillicrap2015}. It performs well in applications such as robotic manipulation \cite{Gu2017} and high precision assembly \cite{Inoue2017}. Many manufacturing applications examples have been identified \cite{Thorsten}. However, popular deep learning applications in manufacturing use limited data analytics \cite{WANG2018144}, for cost prediction \cite{NING2020186}, and quality prediction \cite{Bai}. To the best of our knowledge, no investigation of deep learning applied to real-time manufacturing process control exists. This paper demonstrates how DRL algorithms can learn and then control, as an exemplar, a fiber drawing process.

We apply the DRL framework to a compact fiber drawing system (Fig. \ref{desktop}) \cite{FiberDSCC}. Our primary contribution is the development of a DRL-based control system that:
\begin{itemize}
  \item is trained and tested on a \textbf{real physical fiber drawing process} with \textbf{stochastic} behavior and \textbf{non-linear delayed} dynamics;
  \item \textbf{predictively} regulate the diameter to track \textbf{dynamically varying} reference trajectories;
  \item do \textbf{not} require prior \textbf{analytical or numerical models of the system}.
\end{itemize}

\section{Background}\label{Background}

\subsection{Reinforcement Learning}

Reinforcement learning (RL) is a learning method that trains a software agent to maximize a cumulative measure of reward. The RL agent dynamically interacts with the environment. It receives observations and rewards from the environment, then computes an action based on its policy. The environment is affected by the action, and the new reward that corresponds to the new state of the environment is computed. The cycle is repeated until the task is finished. In this cycle, the action that resulted in high reward is 'reinforced'. The agent learns to prefer the action that is similar to the reinforced action. As a result, the agent is optimized to maximize the expected future reward.

\paragraph{State-action value function $Q(s,a)$}

The state-action value function, also called Q-value, represents the expected future reward when taking a certain action at a certain state, thereafter following the agent's policy,
\begin{equation}
Q^\mu(s_t,a_t) = \mathbb{E}[R_t|S_t=s_t, A_t=a_t],
\end{equation}
where $R_t$ and $\mu$ represent the cumulative future reward and the policy of the agent. The cumulative future reward $R_t$, also called return, may be discounted by some factor $\gamma \in [0,1)$,
\begin{equation}
    R_{t_0} = r_{t_0} + \gamma r_{t_0+1} + \gamma^2 r_{t_0+2} + ... = \sum_{t=t_0}^{end} \gamma^{t-t_0} r_t,
\end{equation}
where $r_t$ is the reward at time $t$. The discount factor $\gamma$ models the notion that a state and an action have decreased relation when the state and reward are farther apart in time.

\paragraph{Bellman Equation}

The Q-value may be estimated using a Monte Carlo method, where the agent tries an entire episode and takes the average return for each state-action pair. However, it has to wait until the end of each episode to learn \cite{sutton2018reinforcement}. The Bellman equation is used to address this inefficiency by bootstrapping the Q-value estimation between consecutive time steps,
\begin{equation}
Q^\mu(s_t, a_t) = \mathbb{E}[r_t + \gamma Q^\mu(s_{t+1}, \mu(s_{t+1}))].
\label{Bellman}
\end{equation}
The error between the left and right side of the equation (\ref{Bellman}) is the temporal difference (TD) error. In the TD method, the agent waits only one time step to perform the learning step.

\begin{figure*}[t]
\centering
\begin{subfigure}{0.3\linewidth}
\centering
\includegraphics[width=\linewidth]{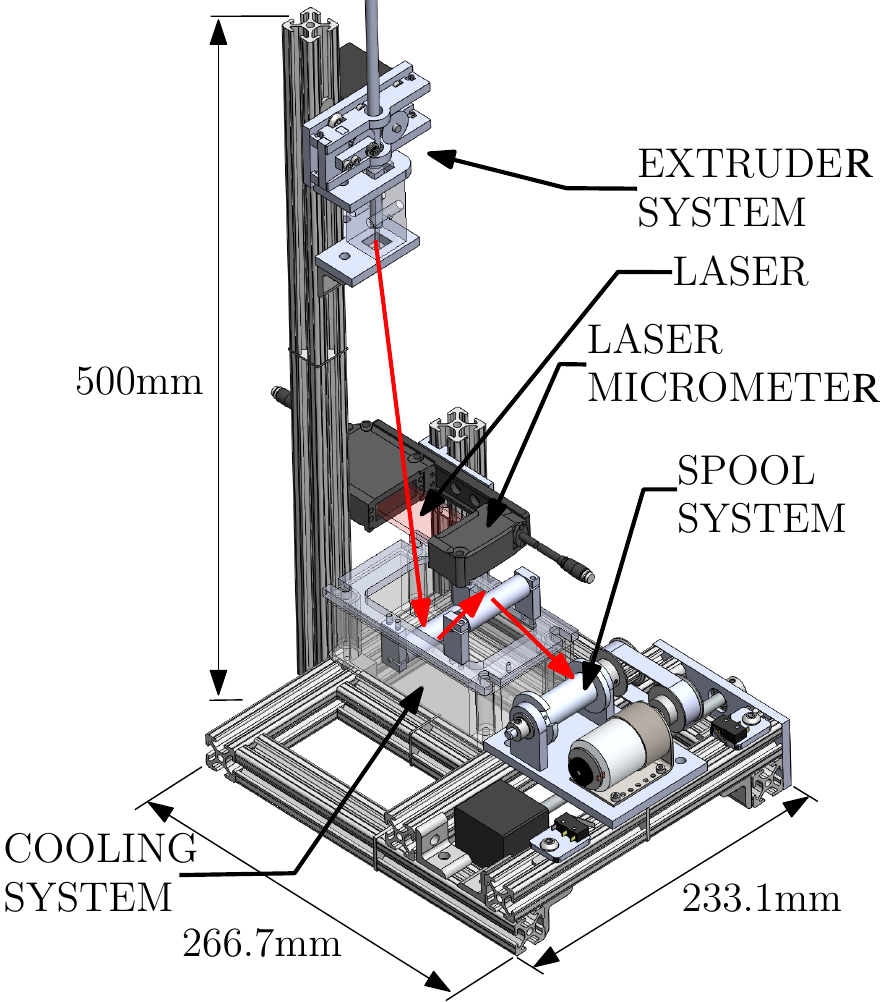}
\caption{Overview} \label{overview}
\end{subfigure}
\begin{subfigure}{0.3\linewidth}
\centering
\includegraphics[width=\linewidth]{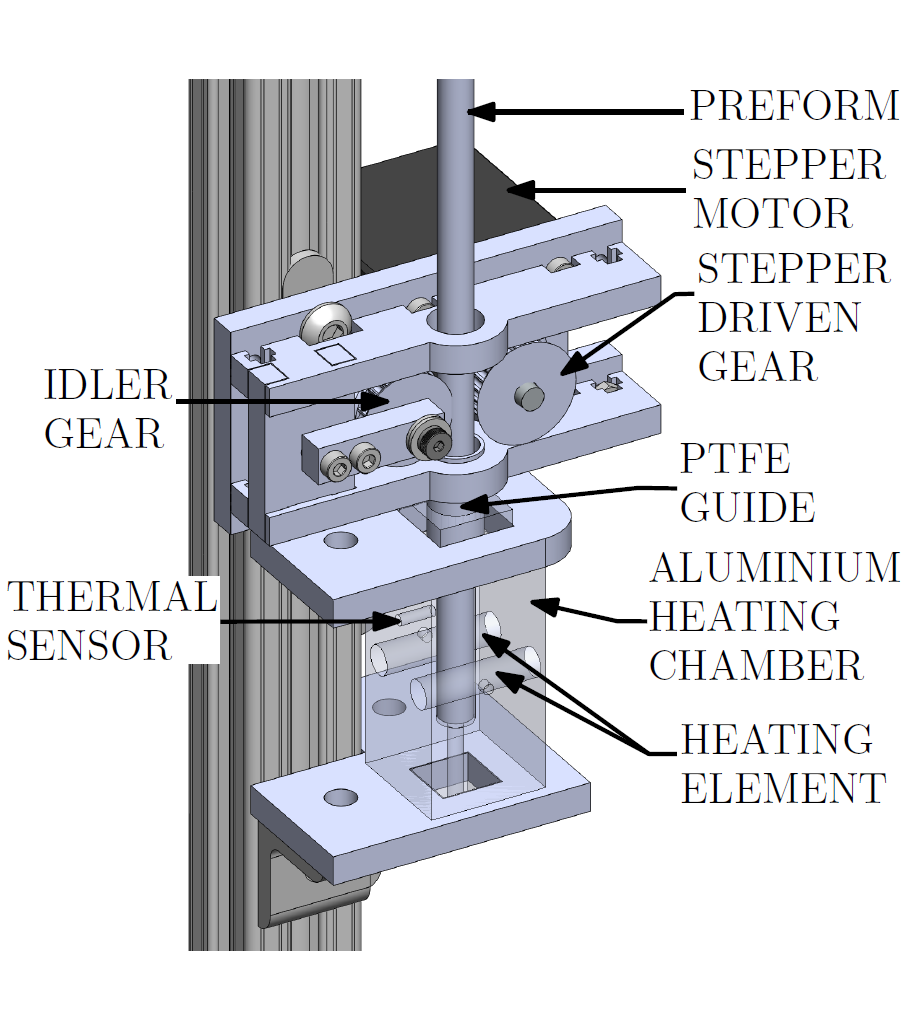}
\caption{The extruder system} \label{extruder}
\end{subfigure}
\begin{subfigure}{0.35\linewidth}
\centering
\includegraphics[width=\linewidth]{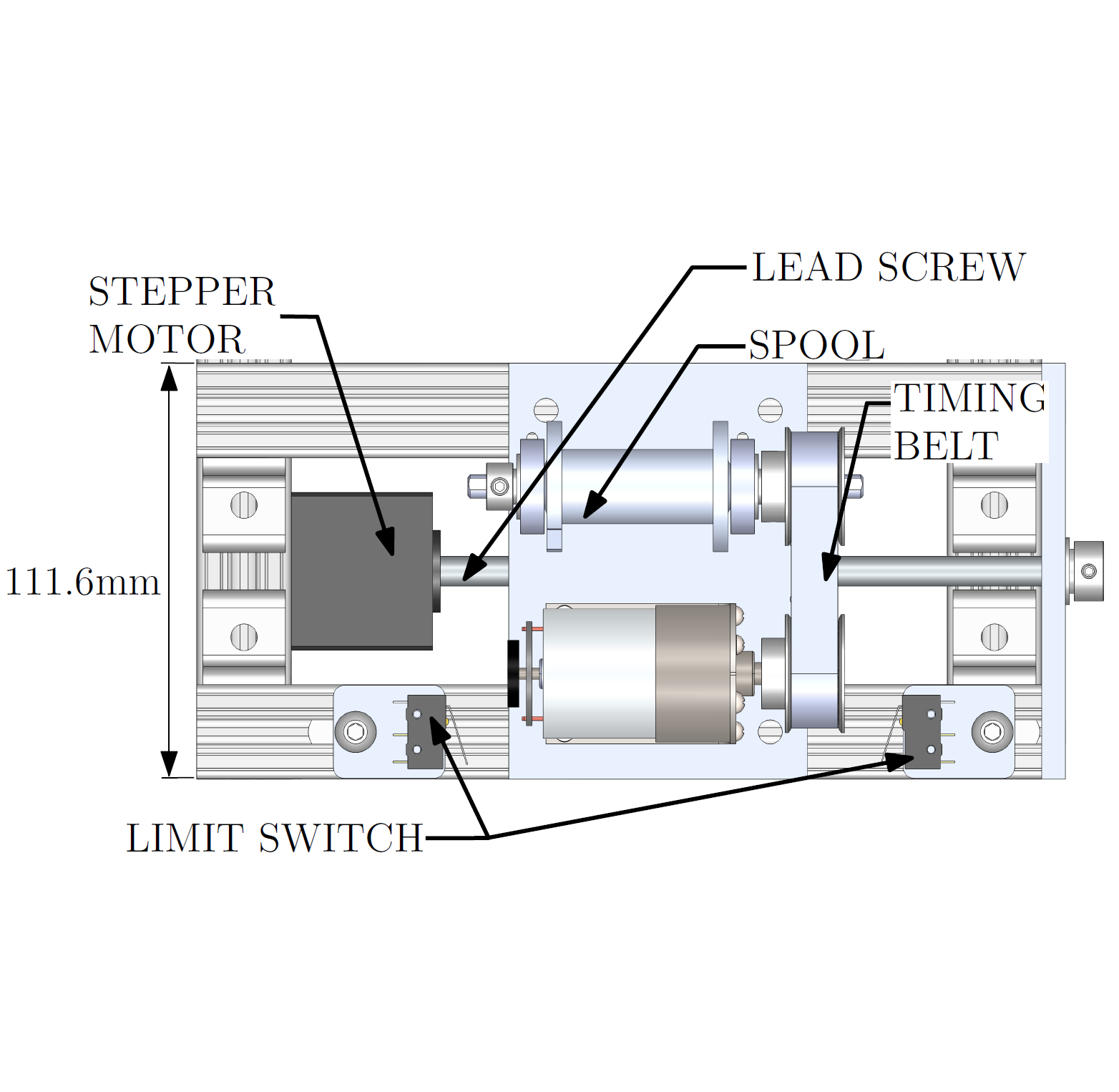}
\caption{The spool system} \label{SPOOL}
\end{subfigure}
\caption{The desktop fiber manufacturing system} \label{system}
\end{figure*}

\subsubsection{Actor-Critic Approach}
The actor-critic approach uses an actor and a critic, each representing the agent (controller) and the Q-function, respectively. The actor observes a state and computes actions. The critic evaluates the actor's action by estimating the Q-value. Based on the critic's evaluation, the actor is updated along the direction that increases the Q-value. Simultaneously, the critic is updated by minimizing the TD error in the Bellman equation. Consequently, the critic converges toward the true Q-value and the actor is optimized to maximize the Q-value. In deep reinforcement learning, multilayer perceptrons are often used as function approximators for the actor and the critic (e.g. deep deterministic policy gradient (DDPG) \cite{Lillicrap2015}). Recurrent Neural Networks (RNN) are used to incorporate the history of the observations and actions (e.g. recurrent deterministic policy gradient (RDPG) \cite{Heess2015}). In some cases, multiple critics are used (e.g. twin delayed DDPG (TD3) \cite{Fujimoto2018}).

\subsubsection{Partial Observability and Long Short Term Memory (LSTM)}
Full observability means that the observations can be used to fully represent the current state. If the state is fully observed, the probability distribution of the next state only depends on the current observation and the current action; this is a Markov decision process (MDP) model.  Partial observability means that the current observations only represent part of the full state. The probability distribution of the next state cannot be determined solely from the current observation and the current action; this is a partially observed Markov decision process (POMDP).

In the fiber drawing process discussed in \Romannum{3}, sensors measure the diameter of the fiber, the temperature of the heating chamber, the speed of the spool motor, and the feed-rate of the extruder. However, this is not a full system observation due to delayed dynamics. The diameter response to the feed-rate or the spool-speed change is delayed by the few seconds it takes for the material to flow through the system. Therefore, the history of observations and actions are needed to predict future states. RNNs provide an approach to this problem. RNNs pass activation values to consecutive time steps so that the input history, from previous time steps, is used when computing new outputs.

A neural network is generally updated along a gradient direction. In a typical feed-forward network, the gradient is backpropagated from the output layer to the input layer. In RNNs, the gradient is also backpropagated through time (BPTT) in order to take into account previous inputs while updating. LSTM is a type of RNN that enables BPTT to reach farther timesteps, preventing the gradient from vanishing by using a gate mechanism \cite{Hochreiter1997}. Therefore, LSTM is widely used in domains such as robot manipulation \cite{Song2018}, self-driving cars \cite{Altche2017}, and language modeling \cite{Sundermeyer2012}.


\section{Mechanical System}\label{MechD}

The compact fiber production system is shown in Fig. \ref{desktop} and \ref{system}. The mechanical design as previously described \cite{FiberDSCC} has been improved.

\subsection{Extrude}

The extruder subsystem is shown in Fig. \ref{extruder}. The extruder subsystem is composed of a heating chamber and a feeding actuator. The heating chamber has a sensor and heating elements to control the temperature. The feeding actuator feeds the preform into the heating chamber at a controlled speed. 

The heating chamber has two cartridge heaters each operating at 40 watts. A resistance temperature detector (RTD) is used. The feeding actuator is composed of a stepper motor and an idler. The feed-rate is controlled by the stepper motor speed.

    \subsection{Cool and Spool}

After the extruder subsystem, the fiber passes through the cooling and spooling subsystems. The general path of the fiber is shown in Fig. \ref{overview} with a red arrow. The fiber passes through the laser micrometer for diameter measurement before it enters the coolant. After the cooling system, the fiber enters the spool system.

The detailed design of the spool system is shown in Fig. \ref{SPOOL}. The main function of the spool system is to collect the fiber and to enable fast-response control of the fiber diameter. As the spool spins faster, the fiber tension increases and the diameter reduces, assuming a fixed preform feed-rate. The spool is rotated by a DC motor with an encoder. The spool and the DC motor are mounted to the stage that is actuated by a lead screw and a stepper motor. The stage movement along the lead screw allows the fiber to be wound evenly along the length of the spool.

\section{Learning Algorithm} \label{LearnAlg}

\begin{figure}[t]
\centering
\includegraphics[width=\linewidth]{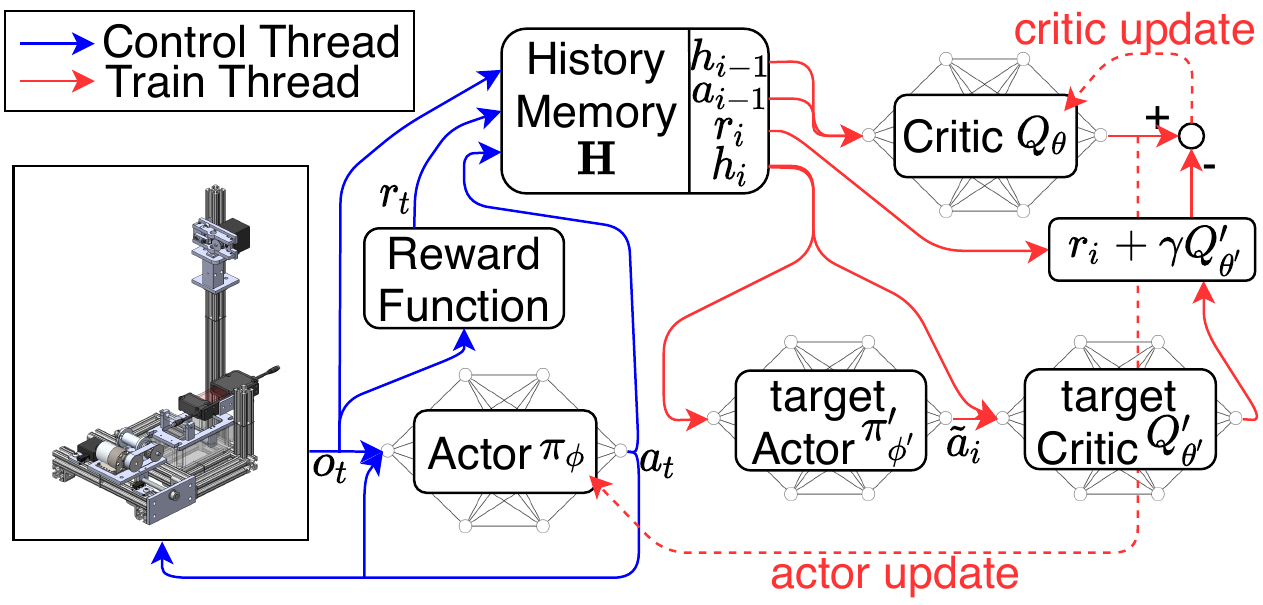}
\caption{Learning algorithm overview}
\label{RLoverview}
\end{figure}

The learning algorithm inspired by \cite{Lillicrap2015,Heess2015, Song2018, Hausknecht2016} is used for training the controller. Fig. \ref{RLoverview} shows an overview of the learning method. Four LSTM networks compose the overall model: actor $\pi_\phi$, critic $Q_\theta$, target actor $\pi'_{\phi'}$ and target critic $Q'_{\theta'}$. The target actor and the target critic are time-delayed copies of the actor and the critic, which facilitates a stable convergence of actor and critic \cite{Lillicrap2015}.  [$\phi$, $\theta$, $\phi'$, $\theta'$] are the parameters of each network. The networks are manipulated in the three sub-processes: an initialization thread, a control thread, and a train thread.

\subsection{Network Structure}

\begin{figure}
\centering
\begin{subfigure}{0.49\linewidth}
\centering
\includegraphics[width=\linewidth]{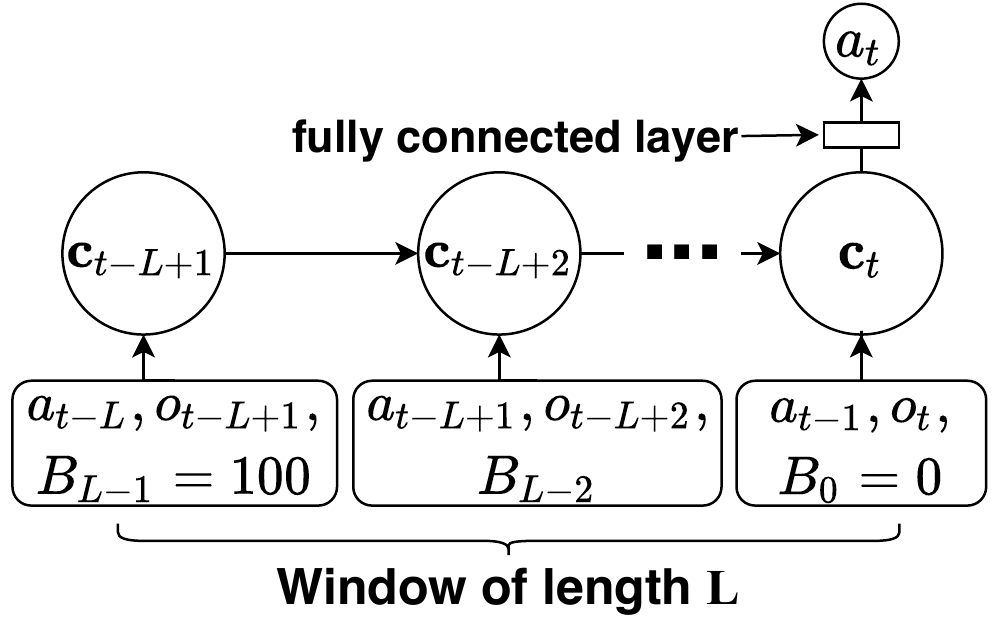}
\caption{Actor network} \label{NETWORK:ACTOR}
\end{subfigure}
\begin{subfigure}{0.49\linewidth}
\centering
\includegraphics[width=\linewidth]{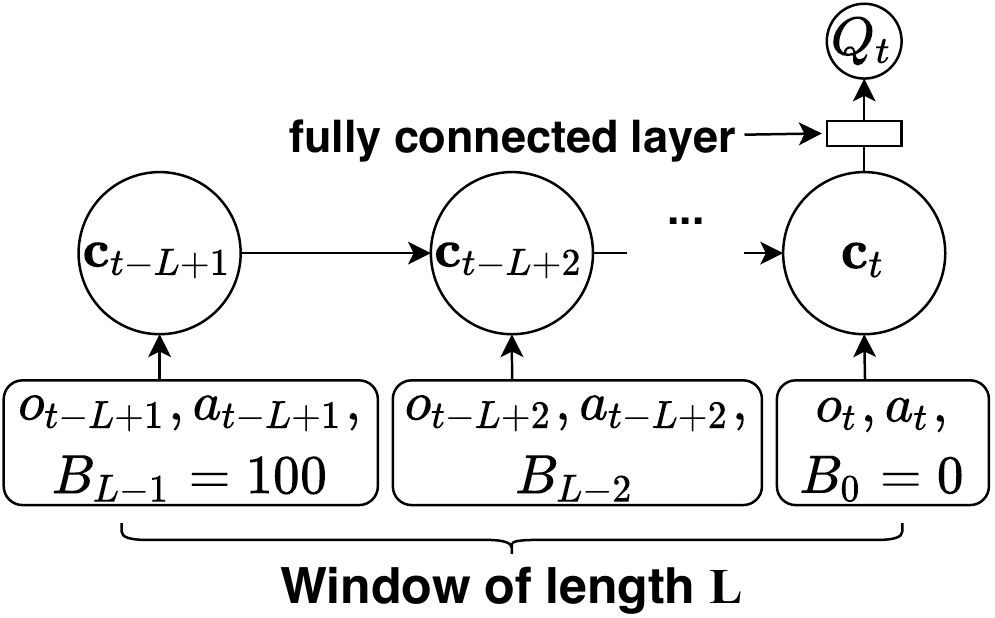}
\caption{Critic network} \label{NETWORK:CRITIC}
\end{subfigure}
\caption{Network structures of actor and critic} \label{Network}
\end{figure}

The network structure of the actor and the critic are shown in Fig. \ref{Network}. The structure of the target actor and the target critic is identical to that of the actor and the critic. Each of the circles ($\textbf{c}_i$) in the figure represents the LSTM network. The number of layers of each network is set to five and the number of nodes in each layer is set to 512. The networks recurse through $L$ timesteps. $L$ is the window length, the span of time during which the networks take inputs. At each timestep, the action taken at one timestep before ($a_{t-1}$) and the following observation ($o_t$) is fed into the actor. An observation at each timestep ($o_t$) and the following action ($a_t$) is fed into the critic. The outputs are computed by passing the activation values of the last recursion through a fully connected layer.

\subsubsection{Observation ($o_t$)}

includes the below components

\begin{itemize}
\item spool's angular speed ($\omega_t$)
\item fiber diameter ($d_t$)
\item summation of the extruder feed-rate ($\sum_{t=0}^{} f_t \Delta t$)
\item current and future reference diameter ($d^{\text{ref}}_t$, $d^{\text{ref}}_{t+10}$, ..., $d^{\text{ref}}_{t+50}$)
\end{itemize}

The spool's angular speed and the fiber diameter are measured by the motor encoder and the laser micrometer. The summation value of the feed-rate represents the cumulative length of fiber produced from the start time of the spool. As the fiber is drawn and wrapped around the spool the effective radius of the spool increases; therefore, the linear speed of the fiber increases over time for a spool with a fixed angular speed. Given this very strong relation between the summation value (cumulative length) and the effective radius of the spool, we include the summation value in the observation.

The last component of the observation is the current and future reference diameter. It is obtained from the desired diameter trajectory $\textbf{d}^{\text{ref}}=\{d^{\text{ref}}_1,d^{\text{ref}}_2,...,d^{\text{ref}}_T\}$, which we want to produce. The reference diameter at the present timestep and also several future timesteps are used ($d^{\text{ref}}_t$, $d^{\text{ref}}_{t+10}$, ..., $d^{\text{ref}}_{t+50}$); the observation looks as far as 50 timesteps (12.5 seconds) ahead. Therefore, the agent predictively controls the system based on the future desired diameters.

\subsubsection{Action ($a_t$)}

\begin{figure}
\centering
\begin{subfigure}{0.49\linewidth}
\centering
\includegraphics[width=\linewidth]{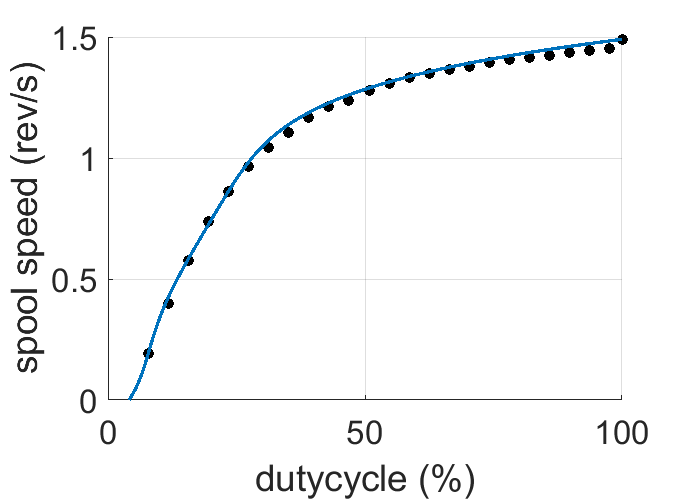}
\caption{Motor's duty cycle vs. speed relation before linear mapping} \label{LinMap:Before}
\end{subfigure}
\begin{subfigure}{0.49\linewidth}
\centering
\includegraphics[width=\linewidth]{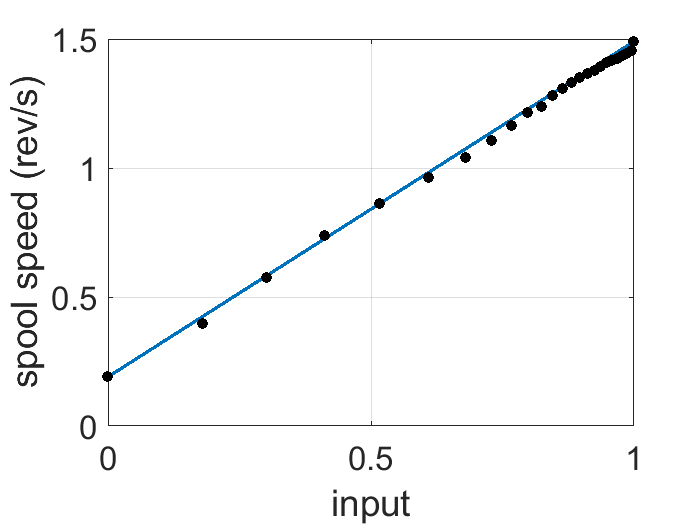}
\caption{Action input vs. speed relation after linearly mapping action input to the speed} \label{LinMap:After}
\end{subfigure}
\caption{Action-speed linear mapping} \label{LinMap}
\end{figure}

includes the below components.

\begin{itemize}
    \item spool input ($a_{\text{sp},t}$)
    \item extruder input ($a_{\text{ex},t}$)
\end{itemize}

The input signaling values for the spool and the extruder have scaled input values between 0 and 100. The spool control input determines the spool motor's PWM duty cycle. The spool input value of 0 and 100 is equivalent to the duty cycle of 7.8\% and 100\%. The extruder input determines the extruder stepper motor's frequency, which is proportional to the feed-rate. The extruder input value of 0 and 100 is equivalent to the feed-rate of 0.09 mm/s and 0.56 mm/s.

The relation between the spool motor's duty cycle and the angular speed measured by an encoder is shown in Fig. \ref{LinMap:Before}. The slope is steep at the low duty cycle and flat at the high duty cycle. Low velocities are very sensitive to the variation of the duty cycle. Consequently, if the spool input ($a_{\text{sp}}$) is mapped just linearly with the duty cycle, then it is hard to precisely control the speed. Therefore, we perform polynomial regression on Fig. \ref{LinMap:Before} and convert the spool input so that it has a linear relation with the speed, as shown in Fig. \ref{LinMap:After}. The extruder input ($a_{\text{ex}}$) is linearly mapped to the stepper motor's frequency because the feed-rate is proportional to the frequency.

\subsubsection{Window Length ($\mathbf{L}$)}

In the original RDPG paper \cite{Heess2015}, the activation values of the LSTM network are propagated from the beginning to the end of each episode. The gradients are back-propagated to the beginning of the episode, and the updates are done between each episode rather than within the episodes. One problem with this method is that computation time increases as the episode gets longer, as the gradient must back-propagate through the entire episode. In the case of the fiber drawing system, each episode is thousands of timesteps long (tens of minutes). Therefore, the computation time is long for each training iteration, reducing the ability to train the model in real-time.

Instead, we consider only the time span that significantly affects the state of the system, rather than the entire episode. We set the length of the time window, through which the networks look into the system (Fig. \ref{Network}). The window size should be long enough to capture the delayed dynamics of the system. The longest delayed dynamic in the fiber drawing system is the delay between a preform feed-rate change and the response in diameter.  It takes approximately 10 seconds (40 timesteps) for a response to appear in the diameter for a step change to the feed-rate. Therefore, the window length should be at least 40 timesteps to capture that delayed dynamic response.

\subsubsection{When-Label ($B_i$)}

To facilitate the learning, when-labels are concatenated as extended inputs. When-labels have scalar values between 0 and 100. They indicate when, in the past, each observation and action occurred. The labels form an arithmetic sequence, where the most recent inputs ($B_0$) have a when-label value of 0 and the oldest inputs ($B_{L-1}$) in the window have a value of 100 (Fig. \ref{Network}). With the when-label, the network incorporates information about when the data was acquired.

\subsection{Initialization}

The parameters of the actor and the critic are initialized using the Glorot initialization approach \cite{Glorot2010}. Then, the parameters are copied to the target actor and the target critic. The empty history memory $\mathbf{H}$ is initialized. Lastly, the recent history buffer $h$ of window length $L$ is initialized. The recent history buffer is a buffer that contains the $L$ most recent observations and actions.

\subsection{Control Thread}

In the control thread, the actor receives an observation from the system and computes an action. First, the actor receives the observation ($o_t$) and the reward ($r_t$) is computed by a reward function. The reward function is defined as the negative value of error between the measured diameter and the reference diameter. Next, the observation ($o_t$) and the previous action ($a_{t-1}$) are appended to the recent history buffer $h$. The actor then takes the recent history buffer as the input and computes a greedy action: $\pi_\phi(h_t)$. An exploration noise ($\epsilon$) is added to the greedy action before the action is executed:
\begin{equation}
    a_t = \pi_\phi(h_t) + \epsilon.
\end{equation}
The exploration noise is required for the actor to consider policies that could be better than the current policy. An Ornstein-Uhlenbeck process \cite{Uhlenbeck1930} with a decay factor $\beta$ is used for the exploration noise. The volatility of the exploration is decreased by the factor of $\beta$ at each timestep. Lastly, the action ($a_t$) is executed on the system as the control input. The reward, observation, and action are added to the history memory $\mathbf{H}$ at each timestep.

\subsection{Train Thread}

The train thread runs in parallel with the control thread. First, N samples of a memory slice, of length $L+2$, are sampled from the history memory $\mathbf{H}$:
\begin{equation}
    (r_{i-L-1},o_{i-L-1},a_{i-L-1},...,r_i,o_i,a_i)
\end{equation}
Next, the target value $y^i$ is computed by the target networks:
\begin{gather}
    h_i \leftarrow (a_{i-L}, o_{i-L+1},..., a_{i-1}, o_i),\\
    \widetilde{a}_i \leftarrow \pi'_{\phi'}(h_i),\;\;
    y_i \leftarrow r_i + \gamma Q'_{\theta'}(h_i,\widetilde{a}_i)
\end{gather}
The target value $y^i$ is used as the right hand side term of (\ref{Bellman}). Then, the loss $J$ (mean squared TD error) of the critic network becomes:
\begin{equation}
J = {1\over N} \sum_{i} (y_i - Q_{\theta}(h_{i-1},a_{i-1}))^2.
\end{equation}
The critic gradient that decreases $J$ can be computed with BPTT:
\begin{equation}
\Delta \theta = {1\over N} \sum_{i} (y_i - Q_{\theta}(h_{i-1},a_{i-1})){\partial Q_{\theta}(h_{i-1},a_{i-1}) \over \partial \theta}.
\end{equation}
By applying the gradient, the critic parameter $\theta$ is updated. The Adam optimizer \cite{Kingma2014} is used as the gradient descent optimizer. After updating the critic, the actor parameter $\phi$ is updated along a gradient that increases the Q-value. The gradient is computed with chain rule,
\begin{equation}
\Delta \phi = {1\over N} \sum_{i} \mathcal{C}\left( {\partial Q_{\theta}(h_{i-1},\pi_\phi(h_{i-1})) \over \partial a}\right) {\partial \pi_\phi(h_{i-1}) \over \partial \phi},
\end{equation}
where $\mathcal{C}(\cdot)$ is a transformation inspired by \cite{Hausknecht2016}, which bounds actions between the maximum and the minimum.
\begin{equation}
\begin{split}
    &\mathcal{C}(\nabla_a) = \\
    &\begin{cases}
    \nabla_a \cdot (a_{max} - a) / (a_{max}-a_{min}),\\
    \qquad \text{if $\nabla_a$ suggests increasing $a$ and $a > a_{max}$}\\
    \nabla_a \cdot (a - a_{min}) / (a_{max} - a_{min})\\
    \qquad \text{if $\nabla_a$ suggests decreasing $a$ and $a < a_{min}$}\\
    \nabla_a, \quad \text{otherwise}\end{cases}
\end{split}
\end{equation}
Lastly, the target actor and the target critic is updated by applying the soft update,
\begin{equation}
(\theta', \phi') \leftarrow (\tau \theta + (1-\tau) \theta', \tau \phi + (1-\tau) \phi'),
\end{equation}
where $\tau$ is a very small positive scalar value. Soft updates of the target networks enable the stable convergence of the model \cite{Lillicrap2015}.


\section{Experiments and Results} \label{Results}

    \subsection{Experimental Setup}
    
    \subsubsection{Hardware System and Hyperparameters}
    
\begin{table}[!t]
\caption{Model hyperparameters}
\label{hyperp}
\centering
\begin{tabular}{|c||c|}
\hline
\textbf{Parameter} & \textbf{Value}\\
\hline
Minibatch size ($N$) & 32\\
\hline
Actor learning rate & 1e-6\\
\hline
Critic learning rate & 5e-6\\
\hline
Soft update factor ($\tau$) & 0.05\\
\hline
History memory ($\mathbf{H}$) size & 75,000\\
\hline
Discount factor ($\gamma$) & 0.99\\
\hline
OU volatility / speed / decay rate ($\beta$) & 10 / 0.1 / 0.999925\\
\hline
window length ($L$) & 50\\
\hline
\end{tabular}
\end{table}
    
The temperature of the heating chamber was set to 80\textdegree{}C, where the fiber drawing is stable with minimal diameter fluctuation. Ethylene-vinyl acetate (Adtech W220-3824 glue-sticks) and room temperature water were used as the material and coolant. Neural network computation was performed on Nvidia's RTX 2080. Sensor measurements and computation results were received and transmitted to PJRC Teensy 3.5 board, an Arduino-based microcontroller. The Teensy 3.5 then controlled the motors and drivers based on the computation results. The timestep was set to 250 ms (4 Hz). The parameters of the algorithms were set to the values in Table \ref{hyperp}.

    \subsubsection{Reward and Training Reference Diameter Trajectory Design}
    
The reward function was defined as,
\begin{equation}
    r_t = - |d_t - d^{\text{ref}}_t| + \alpha f_t + C,
\end{equation}
where $\alpha$ and $C$ are positive scalars, and $d$ and $d^{\text{ref}}$ are in units of one-hundred microns. The first term ($|d_t - d^{\text{ref}}_t|$) represents the error between the reference diameter and the measured diameter at each step. The reward decreases as the error increases. The second term ($\alpha f_t$) is proportional to the feed-rate of the material and thus represents the mass production rate of the fiber. The $\alpha$ was set to 0.106 s/mm, which scales the second term to approximately one-tenth of the first term. This term is needed to ensure the uniqueness of the input action combination. There are two input actions (the spool speed and the extruder feed-rate), which regulate a single output measurement (diameter). Therefore, multiple input action combinations will yield a similar diameter. For example, a combination of a high spool input and a high extruder input can lead to a similar diameter as when a low spool input and a low extruder input is used. However, by adding the second term, the model chooses the combination that maximizes the production rate when there are several other options with similar diameter output. The offset term $C$ was set to 1. If there was no offset term, the reward would be negative at most times. This would lead the model to think that the actions in the operable action range ($a_{\text{min}}$$\sim$$a_{\text{max}}$) are worse than the actions that are outside of the operable range, especially at the early stage of the learning. In this case, the action would be trapped near the operable boundary.

A diameter trajectory that includes random step changes was used for training. Such training trajectories should train the model to track any setpoint step changes.  The duration for each setpoint was 120 timesteps (30 seconds). The setpoint diameter for each step was randomly selected from a uniform distribution between 300 $\mu m$ and 600 $\mu m$. 
  
\begin{figure*}
\centering
\begin{subfigure}{0.495\linewidth}
\centering
\includegraphics[width=\linewidth]{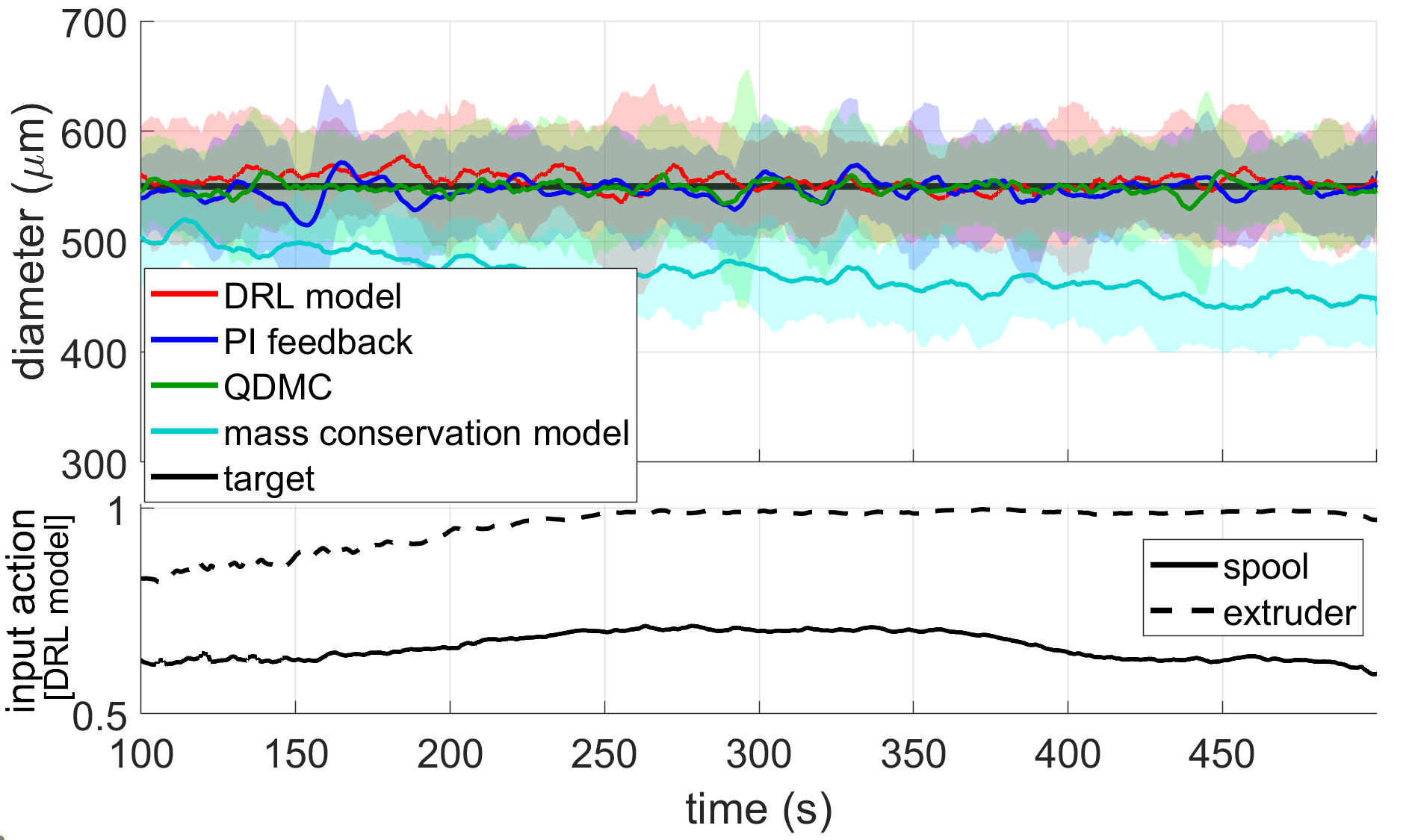}
\caption{Steady trajectory} \label{Test:Steady}
\end{subfigure}
\begin{subfigure}{0.495\linewidth}
\centering
\includegraphics[width=\linewidth]{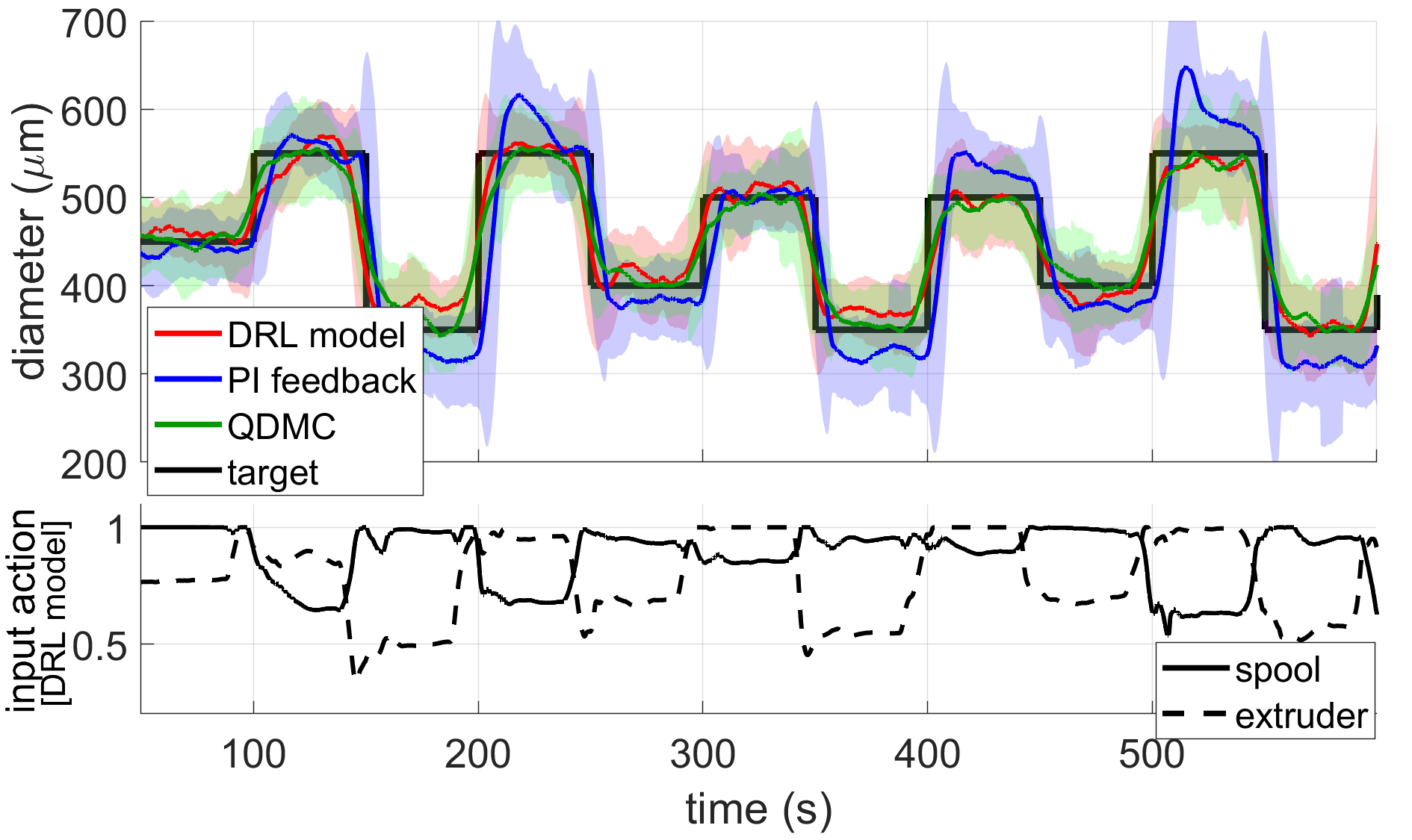}
\caption{Random step trajectory} \label{Test:Step}
\end{subfigure}
\begin{subfigure}{0.495\linewidth}
\centering
\includegraphics[width=\linewidth]{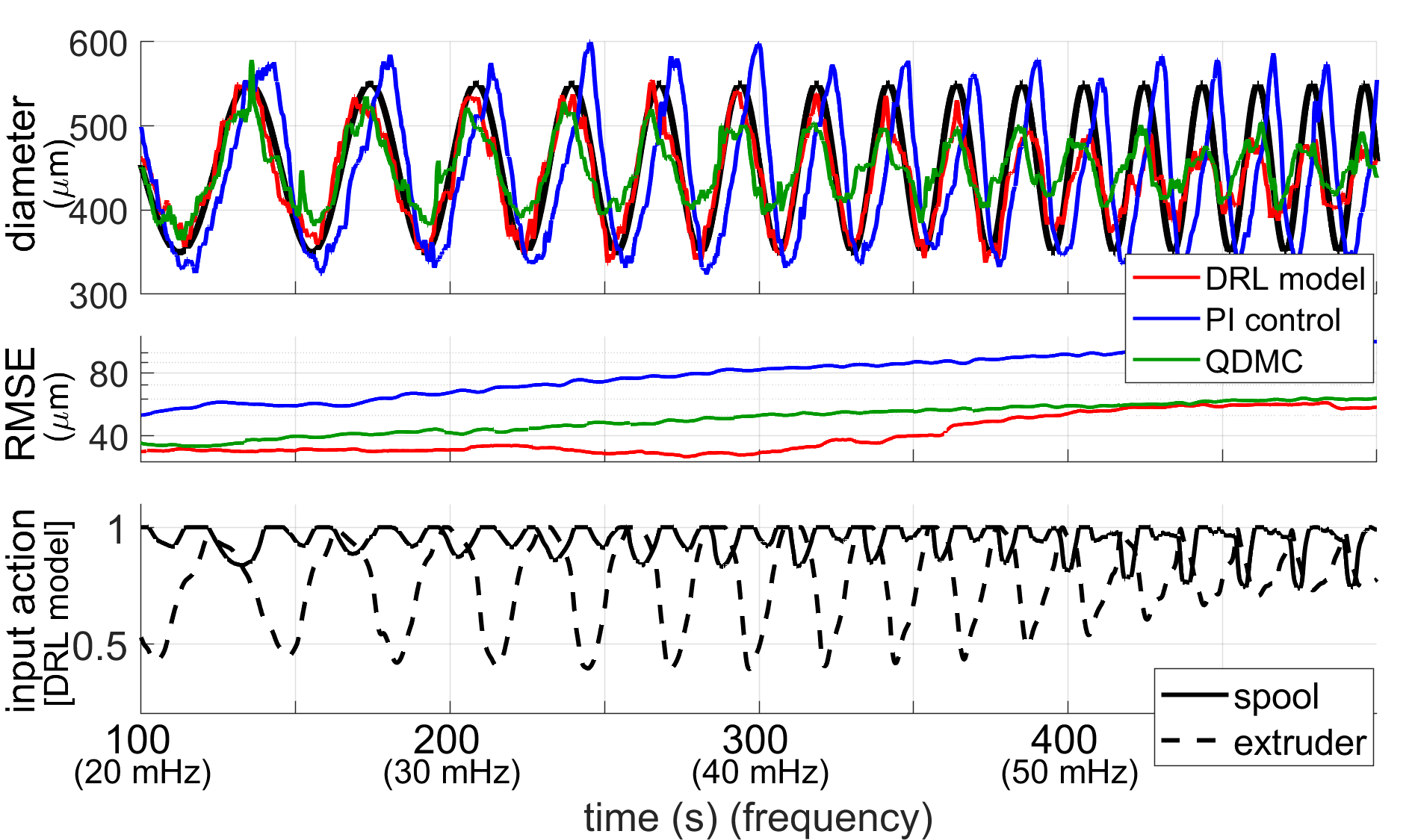}
\caption{Sine sweep trajectory (Amplitude: 100 $\mu m$)} \label{Test:Sin}
\end{subfigure}
\begin{subfigure}{0.495\linewidth}
\centering
\includegraphics[width=\linewidth]{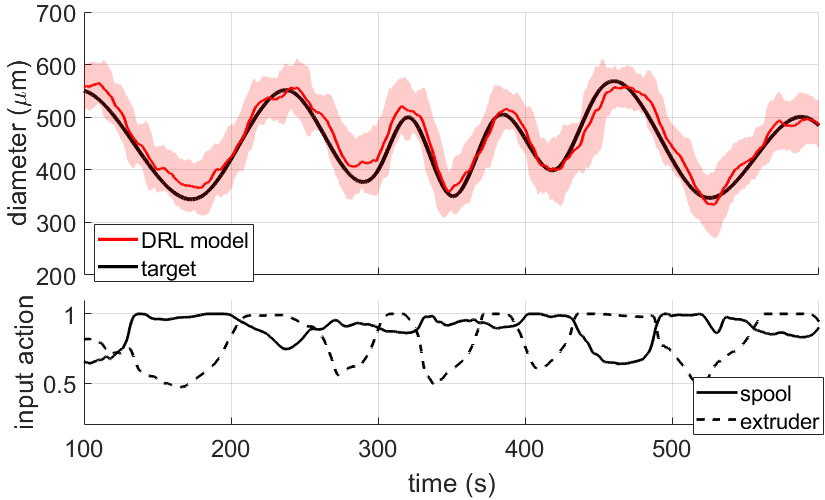}
\caption{Random spline trajectory} \label{Test:Ranspl}
\end{subfigure}
\caption{Target diameter, measured diameter and DRL controller's input actions for each reference trajectory. In (c), RMSE of 500 local timesteps is also plotted.}
\label{Test}
\end{figure*}

\begin{table*}[!t]
\caption{RMSE for each controller and reference trajectory (units are in $\mu m$)}
\label{RMSEtab}
\centering
\begin{tabular}{|c|c|c|c|c|c|c|c|c|} 
\hline
\multirow{2}{*}{\diagbox{Controller}{Ref. traj.}} & \multirow{2}{*}{Steady} & \multirow{2}{*}{Random step} & \multicolumn{3}{l|}{Sine sweep (Amplitude: 100 $\mu m$)} & \multicolumn{3}{l|}{Sine sweep (Amplitude: 50 $\mu m$)}  \\ 
\cline{4-9}
                                                  &                         &                              & 20 mHz & 30 mHz & 40 mHz                         & 20 mHz & 30 mHz & 40 mHz                        \\ 
\hhline{|=========|}
PI                                                & 26.8                    & 61.6                         & 50.3   & 64.5   & 82.7                           & 45.7   & 50.8   & 59.3                          \\ 
\hline
QDMC                                              & 25.5                    & 34.3                         & 36.9   & 41.4   & 50.1                           & 28.3   & 27.4   & 40.5                          \\ 
\hline
DRL                                               & 28.0                    & \textbf{30.4}                         & 33.7   & \textbf{34.4}   & \textbf{33.3}                           & 27.9   & 28.0   & \textbf{31.8}                          \\
\hline
\end{tabular}
\end{table*}
    
    \subsubsection{Baseline Methods}\label{baseline}
    
The performance of the trained model was compared against three baseline methods.

\paragraph{Mass Conservation Open-loop Control}
An open-loop controller, based on the principle of mass flow, assumes that the diameter is set by the rate of the raw material flow:
\begin{equation}
    v_{\text{preform}}A_{\text{preform}} = v_{\text{fiber}}A_{\text{fiber}} = r_{\text{spool}}\omega_{\text{spool}}A_{\text{fiber}},
\end{equation}
where $v$, $A$, $r$, $\omega$ are linear speed, cross-sectional area, radius and angular speed. This model assumes a constant $r_{\text{spool}}$, which means that it does not consider the increase of the effective spool radius due to the fiber stacking on the spool. 

\paragraph{PI Control}
The PI controller regulates the diameter by feedback of the diameter error. The material feed-rate was fixed to 0.37 mm/s and only the spool-speed was controlled with PI control. P and I parameters were optimized using modified particle swarm optimization \cite{shi1998modified,qi2019tuning}. Particle number and maximum iteration were set to 10 and 15, respectively. Each particle was tested on a reference trajectory with three predefined representative step changes: 450 to 550 $\mu m$, 550 to 350 $\mu m$, and 350 to 450 $\mu m$. RMSE was used as the objective function.

\paragraph{Quadratic Dynamic Matrix Control (QDMC)} The model-based QDMC controller uses the step response model of the system \cite{Cutler1980, Carlos1986}. Under the assumptions that the system is linear and time-invariant, it predicts the future diameter and optimizes the present and future inputs by minimizing the cost function:
\begin{equation}
    J = \sum_{i=1}^p (d_{t+i}^{\text{ref}}-\hat{d}_{t+i})^2 + r \sum_{i=0}^{c-1} \norm{\Delta\mathbf{u}_{t+i}}^2,
\end{equation}
where $\hat{d}$ is the predicted diameter in 100 $\mu m$. $\Delta\mathbf{u}_{t+i}$ is the input change. $p$ and $c$ are the prediction and control horizon. $p$ is set to 50 because our DRL controller looks 50 timesteps ahead as explained in \ref{LearnAlg}. $c$ is set 25, half of the prediction horizon. $r$ is a weighting factor that defines ratio of importance between output error and input change. The controller code was developed based on \cite{Adrian}.

The model requires the response in diameter to a step change of each input. The square root of the extruder feed-rate ($\sqrt{f}$) and reciprocal of the square root of command spool-speed ($1/\sqrt{\omega}$) were used as the inputs since the diameter is proportional to $\sqrt{f / \omega}$ according to the mass conservation principle. Each input is normalized such that the minimum and maximum are 0 and 1. The diameter response to the step change of $\sqrt{f}$ was measured at spool-speeds of 0.6, 1.0, and 1.4 revolution/second, then the average response was used for the step response model. For the diameter response to the step change of $1/\sqrt{\omega}$, the average response at extruder feed-rates of 0.19, 0.37, and 0.56 mm/s were used for the model.

The weighting factor $r$ was tuned. If it is too large, the input change is slow and results in a slow diameter response. If too small, the response to input change is too sensitive to disturbances or model error and results in excessive fluctuations in diameter. Weighting factors of 5, 10, 20, 40, 80, 160, and 320 were tested on the same reference diameter trajectory that was used for the training of the DRL controller. The mean error increased significantly at a weighting factor of 5 and 320. Between 10 and 160, the mean error difference was less than 10\%. Therefore, $r$ was set to 40.
\vspace{-12pt}

    \subsection{Experimental Result}

    \subsubsection{Test on Various Reference Trajectories}
    
The DRL controller was trained for approximately 50,000 timesteps (3.5 hours), then tested on several reference trajectories: steady, random step, sine sweep, and random spline. The diameter trajectories and input actions of the DRL controller are plotted in Fig. \ref{Test} and RMSE for each controller and reference trajectory are listed in Table \ref{RMSEtab}. Each controller was tested 5 times for each of the trajectories and the average responses are shown in the plot. In Fig. \ref{Test:Steady}, \ref{Test:Step}, and \ref{Test:Ranspl}, moving average of 40 timesteps is applied and moving standard deviation ($\times$1.96) is shown as the shaded areas.

    \paragraph{Steady Trajectory}
    
\begin{figure*}[t]
\centering
\begin{subfigure}{0.32\linewidth}
\centering
\includegraphics[width=\linewidth]{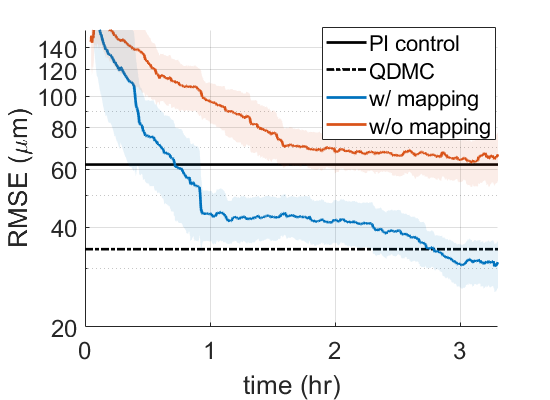}
\caption{Action-speed linear mapping} \label{Lcurve:linmap}
\end{subfigure}
\begin{subfigure}{0.32\linewidth}
\centering
\includegraphics[width=\linewidth]{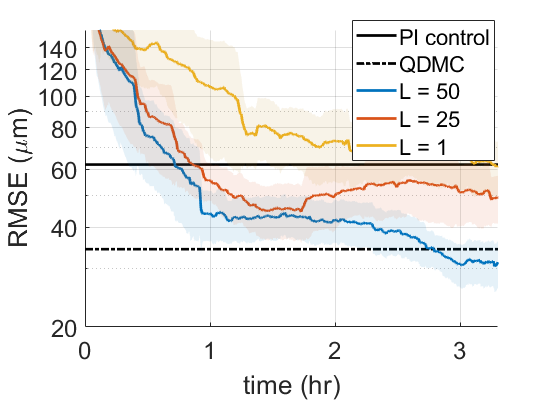}
\caption{Window length} \label{Lcurve:winlen}
\end{subfigure}
\begin{subfigure}{0.32\linewidth}
\centering
\includegraphics[width=\linewidth]{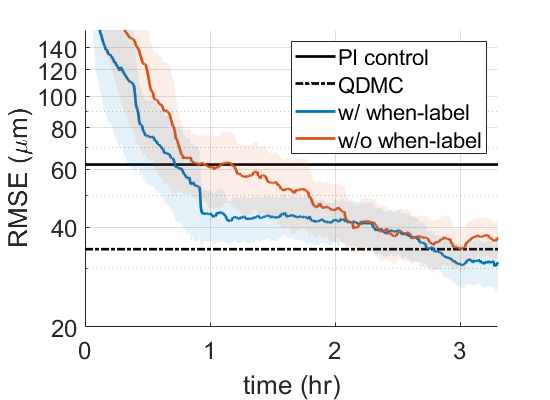}
\caption{When-label} \label{Lcurve:whenl}
\end{subfigure}
\caption{Learning curve comparison. Moving average of 10,000 timesteps ($\sim$0.7 hr) is applied.} \label{Lcurve}
\end{figure*}    
    
\begin{figure*}[t]
\centering
\begin{subfigure}{0.32\linewidth}
\centering
\includegraphics[width=\linewidth]{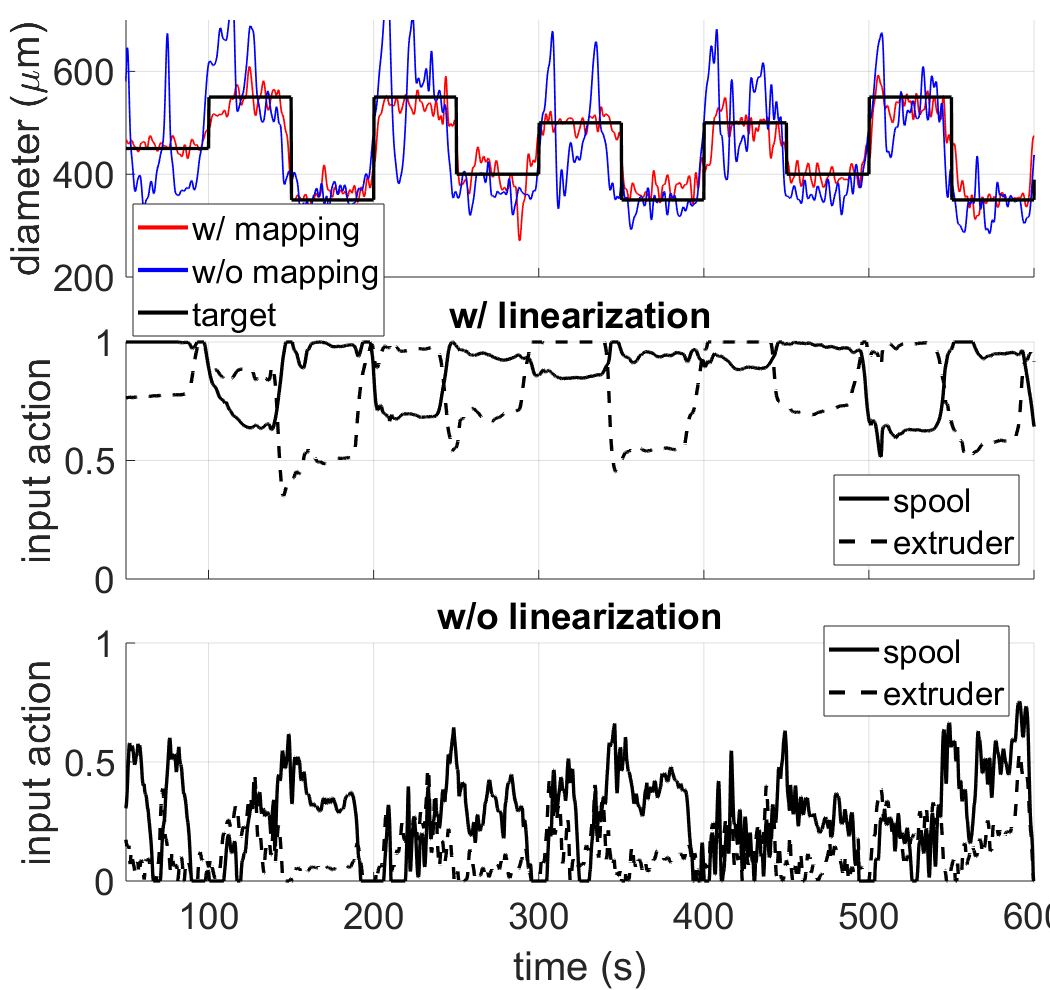}
\caption{Action-speed linear mapping} \label{Comp:linmap}
\end{subfigure}
\begin{subfigure}{0.32\linewidth}
\centering
\includegraphics[width=\linewidth]{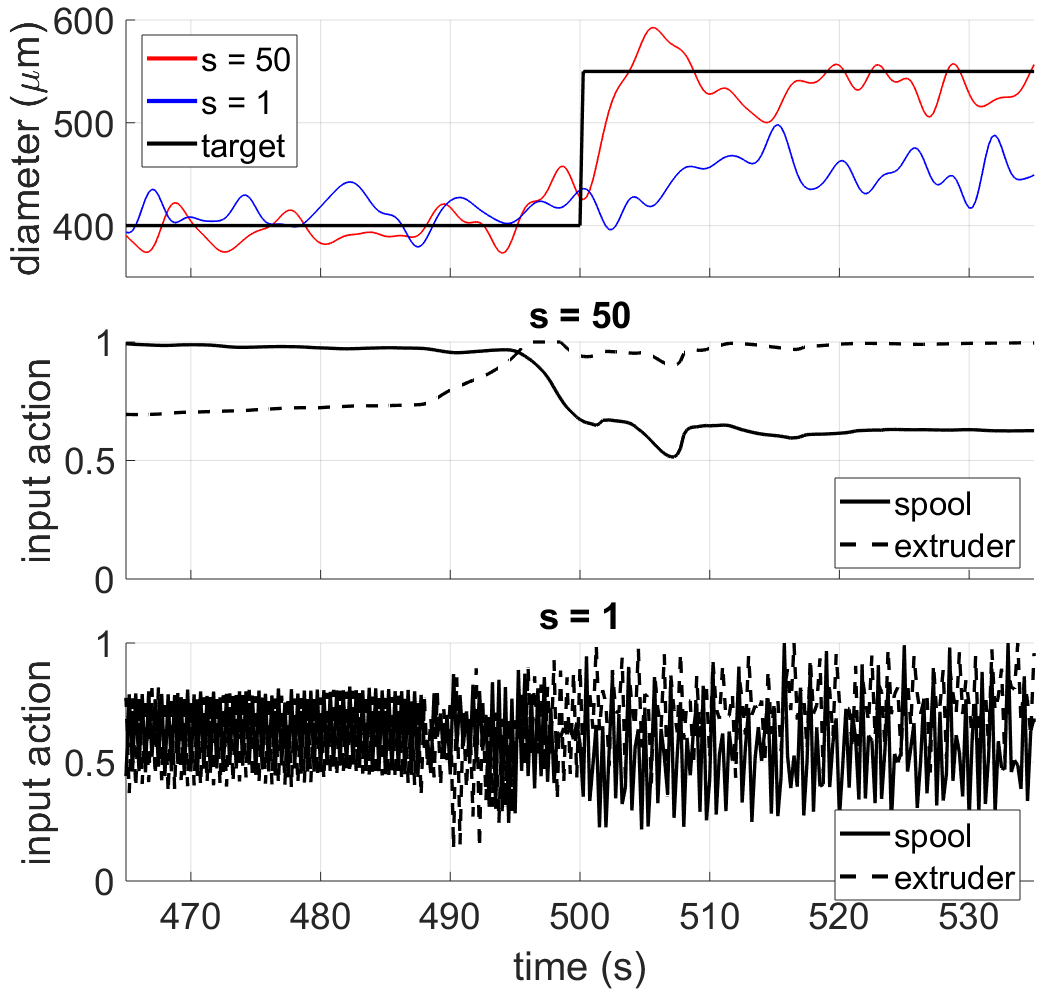}
\caption{Window length} \label{Comp:winlen}
\end{subfigure}
\begin{subfigure}{0.32\linewidth}
\centering
\includegraphics[width=\linewidth]{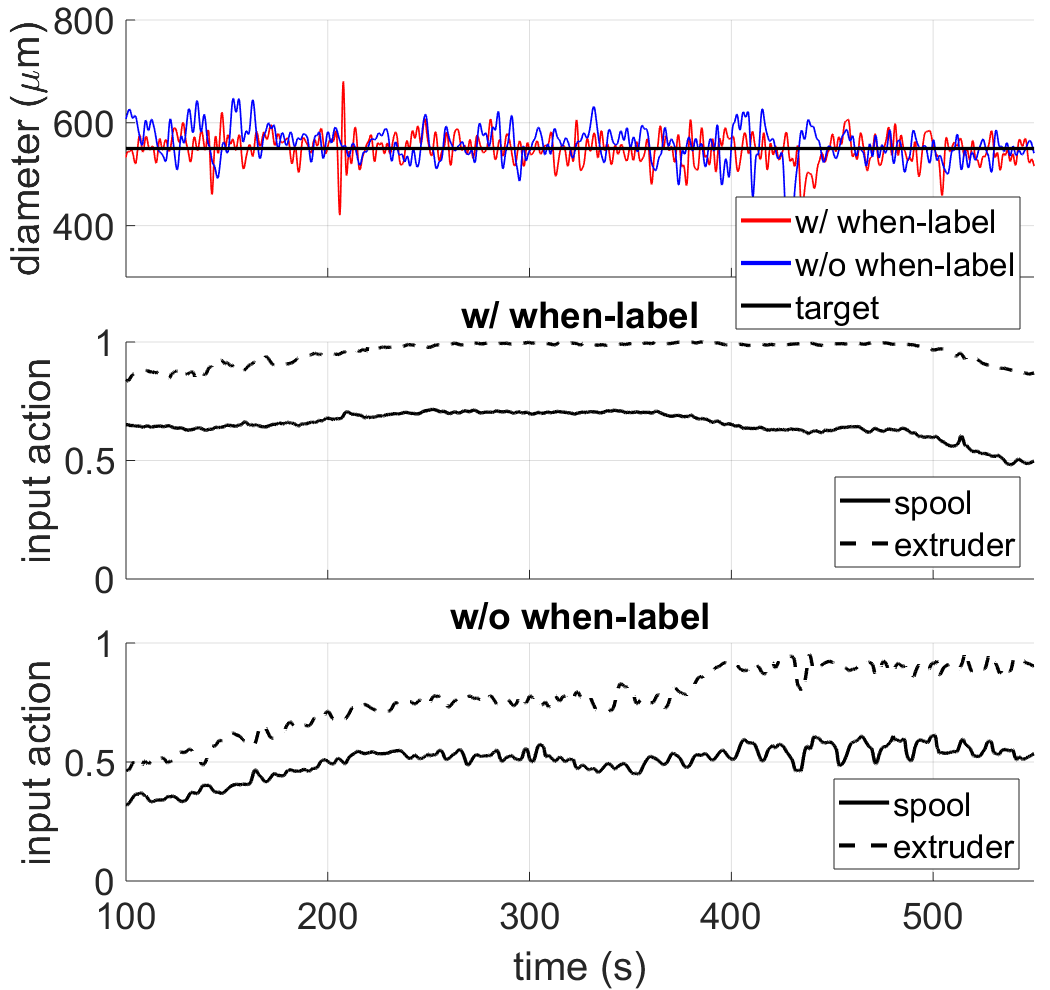}
\caption{When-label} \label{Comp:wlabel}
\end{subfigure}
\caption{Diameter tracking and input action comparison} \label{Comp}
\end{figure*} 

Each controller was tested with a steady reference trajectory at a setpoint of 550 $\mu m$ (Fig. \ref{Test:Steady}). For the mass conservation model, as expected, there was a decreasing trend of diameter with respect to time. As discussed in section \ref{baseline}, the simple model does not consider the increase in the spool radius and maintained the constant angular speed, the linear speed of the fiber increased and the diameter decreased with respect to time. In comparison, the DRL control, PI control, and QDMC maintained diameter close to the reference. In DRL control, the ratio of the extruder input (material feed-rate) to the spool input increased with respect to time. In this way, the effect of the stacking spool is compensated by feeding more material and rotating the spool slower. The DRL control, PI control, and QDMC showed average RMSE of 28.0, 26.8, and 25.5 $\mu m$, respectively. Based on the unpaired t-test, the difference in RMSE is not statistically significant, with 95$\%$ level of confidence.

    \paragraph{Random Step Trajectory}

The random step reference trajectory used for testing had an interval of 50 seconds (Fig \ref{Test:Step}). When the PI controller was used for this trajectory, the measured diameter response showed 3.5 seconds of average time lag as estimated by cross-correlation analysis. The system consistently settled to the target diameter within a single interval and sometimes showed underdamped overshoot. In contrast, the DRL controller and QDMC only showed -0.5 seconds and 0.5 seconds of time lag, respectively. They manipulated input actions with the advance knowledge of the coming step changes. For the DRL controller, the spool input changed 4.5 seconds ahead of the step change and the extruder input changed 8.0 seconds in advance of the step change, both estimated by cross-correlation analysis.

    \paragraph{Continuous Trajectory Trajectory}
    
The DRL controller was trained using a discontinuous step-changing reference trajectory. The response to continuous reference trajectories, a sine sweep and a random spline (Fig. \ref{Test:Sin}, \ref{Test:Ranspl}), was evaluated. The sine sweep trajectory swept from 0.01 Hz to 0.06 Hz with a sweeping rate of $10^{-4}$ Hz/s. The mean was 450 $\mu m$ and two amplitudes, 100 and 50 $\mu m$, were tested. The random spline reference was generated by connecting several points with a B-spline curve. The diameters of each spline points were set between 350 $\mu m$ and 550 $\mu m$ and the timestep difference between adjacent points were set between 20 timesteps (5 seconds) and 80 timesteps (20 seconds), introducing multiple frequency components with multiple amplitudes.

In the sine sweep trajectory (Fig. \ref{Test:Sin}), all of the controllers showed an increasing trend in RMSE as the sine frequency increased because there is a physical limit on how fast the system can respond to the input changes. The PI controller showed significantly larger RMSE than other methods at all frequency range. For the reference trajectory with 100 $\mu m$ amplitude, the DRL controller and QDMC showed similar RMSE at below 20 mHz. Between 20 mHz and 50 mHz, the DRL controller showed less RMSE than QDMC. The DRL controller regulated the RMSE to under 40 $\mu m$ until the frequency reached 45 mHz, while QDMC was able to regulate only until 25 mHz. The RMSE difference between the DRL controller and QDMC was smaller when the amplitude was 50 $\mu m$ (Table \ref{RMSEtab}). The DRL controller also tracked the random spline trajectory by gradually varying the input actions (Fig. \ref{Test:Ranspl}).

\section{Discussion} \label{Disc}

\subsection{Comparison with PI Control and QDMC}

The PI controller only considers the present and past error, so it is not predictive. It results in a significant time lag when the system involves delayed dynamics. Contrastly, the DRL model predictively controls the diameter by using the future reference diameter as the input to the model.

The PI and QDMC controllers require careful tuning to ensure the optimal performance, of the P and I parameters and $r$ parameter respectively. We conducted 150 experiments to tune the PI controller. In the DRL model, we do not tune the parameters in advance with multiple experiments; the model learns the optimal policy during operation.

The PI controller was tuned at specific setpoints; it does not ensure the optimal performance at other setpoints, especially in a non-linear system. Similarly, QDMC was tuned with a specific step response model and assumes a linear and time-invariant (LTI) system. However, our fiber drawing system is not linear. For example, the diameter response to the step change in spool velocity is asymmetric depending on whether the velocity changes from low to high or high to low. Also, the system is not time-invariant because the produced fiber increases the effective radius of the spool as it is wrapped around the spool. Therefore, errors occur in the QDMC's prediction and results in suboptimal performance.  This is not a critical problem when tracking a steady reference trajectory or responding to a step change in desired diameter with long step interval.  It is critical when the target diameter fluctuates rapidly as in Fig. \ref{Test:Sin}. On the other hand, the DRL controller uses a neural network (NN). A NN can approximate non-linear functions. Accordingly, the DRL controller learned how to deal with the nonlinearity of the system without prior analytical or numerical models.

The results with continuous reference trajectories suggest that the learned DRL controller can be used to track trajectories similar to those used in the training process, and used to track trajectories not faced during the training process. Such generalization is enabled by using NN as a non-linear function approximator for the actor and critic. According to Lillicrap \cite{Lillicrap2015}, the use of such approximators is essential for generalizing to large state spaces.

\subsection{Ablative Analysis}

    \subsubsection{Effect of Action-Speed Linear Mapping}
    
Fig. \ref{Lcurve:linmap} shows that the action-speed linear mapping is critical to achieving good performance. The model without the linear mapping converged to the average reward approximately 0.2 smaller than the model with the mapping. This means that the average diameter error was approximately 20 $\mu m$ bigger. The model showed poor performance especially when the reference diameter was large, where low spool-speed is required (Fig. \ref{Comp:linmap}). This is because it is hard to control the speed precisely at the low speed range if action-speed is not linearly mapped. Linearly mapping the spool action to the speed enables the model to control the speed precisely throughout the entire speed range and result in better performance.

    \subsubsection{Effect of Window Length}
    
Models with several different window lengths are compared. The learning curve comparison shows that the window length must be long enough to achieve optimal performance (Fig. \ref{Lcurve:winlen}). When the window length is 1, it computes the input action based on only one timestep of observation. Therefore, it does not use the previous history of the process. Also, it cannot capture the stochastic nature of the system. As a result, the computed input action fluctuates violently as shown in Fig. \ref{Comp:winlen}. The model with window length 25 was also not as good as that with window length 50. This is because 25 timesteps (6.25 seconds) are not enough to capture the delayed dynamics when the step change occurs. As described previously, change in the extruder input should occur 8.0 seconds earlier than the diameter step change. Therefore, the window length should be at least 32 timesteps (8.0 seconds) to capture these delayed dynamics.

    \subsubsection{Effect of When-label}
    
Fig. \ref{Lcurve:whenl} shows that the when-label accelerates the learning, especially at the beginning. The when-label helps the learning of the model by providing additional information about when the data was observed. Thereby, the model can learn the dynamics of the process faster than when the label is not provided. Also, the model with the when-label computed more consistent outputs. In comparison, the model without when-label showed some fluctuation in its outputs, as shown in Fig. \ref{Comp:wlabel}. This high-frequency fluctuation is unnecessary since the system cannot physically respond at those rates.

\section{Conclusion} \label{Conc}

We introduced the compact fiber drawing system and controlled it with a DRL-based strategy. We regulated the fiber diameter to track time-varying reference trajectories. By customizing DRL algorithms, we optimized the performance of the system in terms of tracking error. Without analytical or numerical models of the physical system, the controller learned to track various types of reference trajectories under the stochasticity and the non-linear delayed dynamics of the system. The DRL controller was generalized to track diameter trajectories that it had not experienced in the training process. Using the same framework, a pretraining step can be added, where the controller learns from pre-existing production data and reduces the time required for the online training.

\section*{Acknowledgment}

We thank the MIT Skoltech Initiative for the partial funding that allowed us to prototype the fiber system. We also thank the MIT Professional Education Office for collaboration in developing the Smart Manufacturing Leadership Program which gave us the opportunity to test and debug our system.

\ifCLASSOPTIONcaptionsoff
  \newpage
\fi

\bibliography{IEEEabrv,FiberDK}
\bibliographystyle{IEEEtran}

\begin{IEEEbiography}[{\includegraphics[width=1in,height=1.25in,clip]{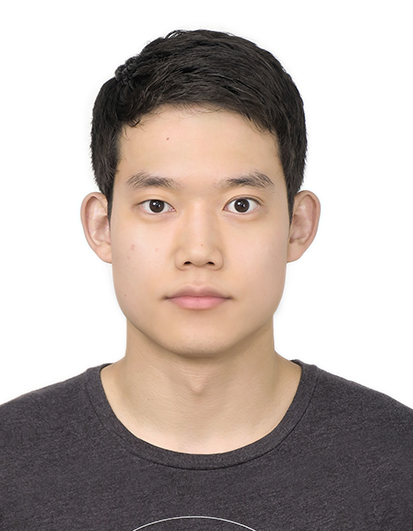}}]{Sangwoon Kim}
is currently attending MIT as a graduate student in the Department of Mechanical Engineering. He received a Bachelor’s degree in Mechanical and Aerospace Engineering from Seoul National University and a Master's degree in Mechanical Engineering from MIT. He is interested in applying machine learning algorithms to physical environments such as manufacturing processes.

\end{IEEEbiography}

\begin{IEEEbiography}[{\includegraphics[width=1in,height=1.25in,clip]{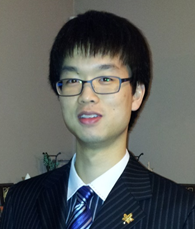}}]{David Donghyun Kim}
is a research specialist at MIT. He received a Bachelor of Applied Science in Mechanical Engineering with Management Science Option from University of Waterloo and a PhD in Mechanical Engineering from MIT. He is interested in mechanical design for robotic systems. His fundamental research background in CAD/CAM especially focusing on 5-axis CNC milling allowed him to design with manufacturing in mind.
\end{IEEEbiography}

\begin{IEEEbiography}[{\includegraphics[width=1in,height=1.25in,clip,keepaspectratio]{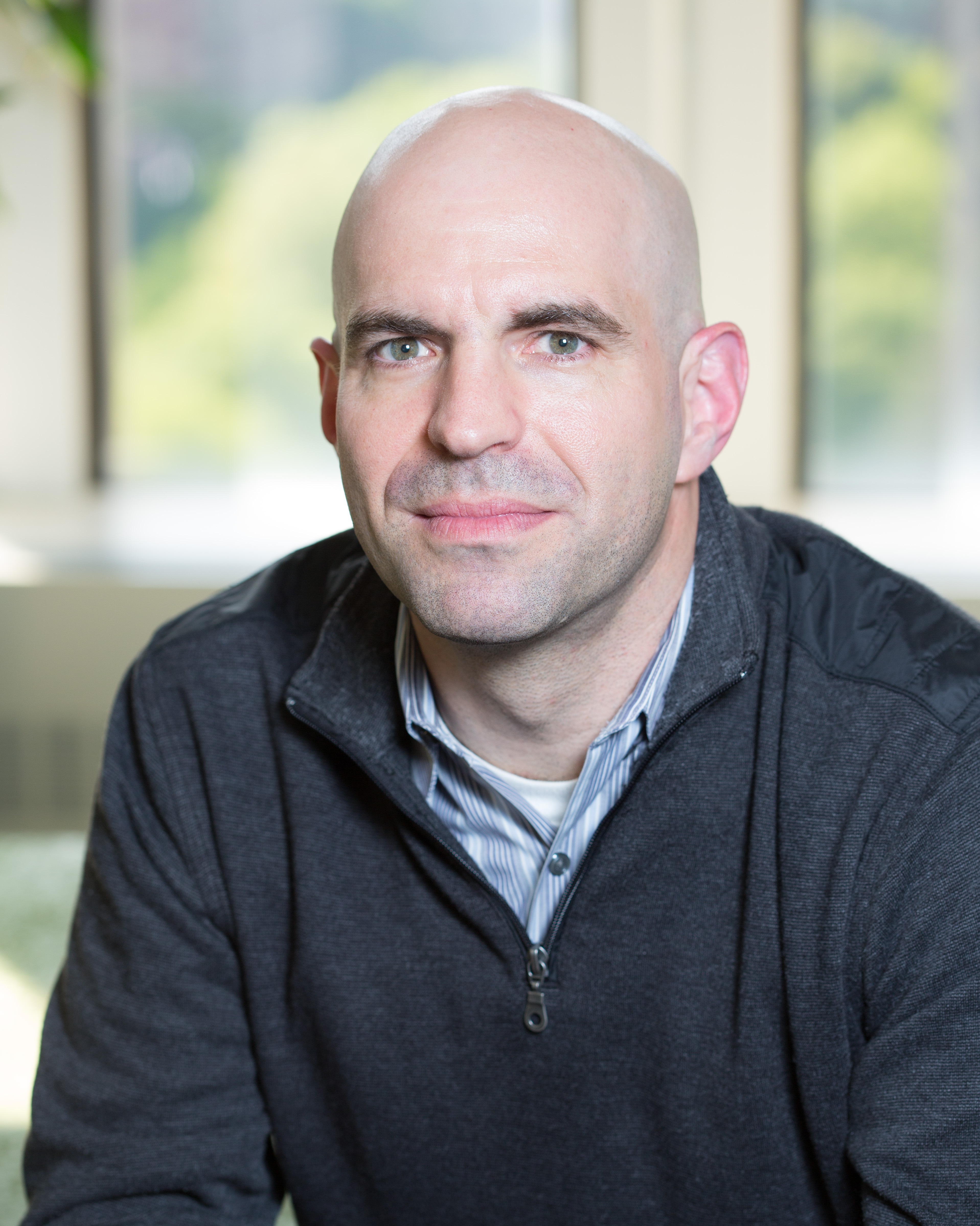}}]{Brian W. Anthony}
is currently Associate Director of MIT.nano. He has extensive experience in market driven technology innovation, product realization, and business entrepreneurship and commercialization at the intersection between information technology and advanced manufacturing. His research and product development interests cross the boundaries of manufacturing and design, medical imaging, computer vision, acoustic and ultrasonic imaging, large‐scale computation and simulation, optimization, metrology, autonomous systems, and robotics.
\end{IEEEbiography}

\end{document}